\documentclass[12pt]{iopart}
\usepackage{iopams}  
\usepackage{graphicx}

\def \beq {\begin{equation}}
\def \edq {\end{equation}}
\def \bes {\begin{subequations}}
\def \eds {\end{subequations}}
\def \calh {{\cal{H}}}
\def \calg {{\cal{G}}}
\def \bcalg {{\mathbf{\cal{G}}}}
\def \veps {\varepsilon}

\def \bt {\mathbf{t}}
\def \bGamma {\mathbf{\Gamma}}
\def \bSigma {\mathbf{\Sigma}}

\begin{document}

\title[Transport through Majorana nanowires]{
Transport through Majorana nanowires attached to normal leads}
 
\author{Jong Soo Lim$^1$, Rosa L\'opez$^{1,2}$ and Lloren\c{c} Serra$^{1,2}$}

\address{$^1$ Institut de F´\'{\i}sica Interdisciplin\`aria i Sistemes Complexos IFISC (CSIC-UIB),
E-07122 Palma de Mallorca, Spain\\
$^2$ Departament de F\'{\i}sica, Universitat de les Illes Balears, 
E-07122 Palma de Mallorca, Spain
}
\ead{llorens.serra@uib.es}
\begin{abstract}
This paper presents a coupled channel model for transport in 2D 
semiconductor Majorana nanowires coupled to normal leads.
When the nanowire hosts a zero mode, conspicuous signatures on the 
linear conductance are predicted. 
An effective model in second quantization allowing 
a fully analytical solution is used to clarify the physics.
We also discuss the nonlinear current
response ($dI/dV$)
\end{abstract}

\submitto{\NJP}

\maketitle

\section{Introduction}

The theoretical proposal that Majorana modes exist in semiconductor devices
\cite{sau10,ali10,ore10}
and their  subsequent detection in InSb wires
\cite{mou12,den12,rok12}
has opened up a new subfield of research on nanostructure properties
(see Refs.\ \cite{wil09,fra10,bee11,ali12} for reviews). It was originally proposed 
by Majorana
that a massless elementary particle, called a Majorana particle,
could exist with the peculiar property 
of being its own antiparticle.
Similarly, a Majorana mode of a semiconductor nanowire is a zero-energy state
that remains invariant after charge conjugation.
These states are 
quasiparticle excitations
localized on the tips of a finite but long enough
wire and they are well separated from the rest of the spectrum of eigenstates
by an energy gap.

The existence of Majorana modes in a semiconductor wire requires the 
presence of the following physical ingredients: a) Zeeman coupling
between spin and magnetic field, b) Rashba spin-orbit
interaction and c) superconductivity \cite{nad10,nil09,nil12}. 
The latter can be induced by proximity with a superconductor material 
and it introduces 
the concept of electron-hole symmetry \cite{fu08}.  
The Rashba spin-orbit interaction is a relativistic effect originating in 
the quantum well asymmetry
in the perpendicular direction to the nanostructure plane. 
In the present context this interaction introduces chirality by connecting 
the state of motion with spin. 
The Zeeman coupling in
semiconductors like InAs and InSb is 
quite large even for relatively low magnetic fields
due to the large  g factors of these materials. It breaks
Kramers degeneracy since the system is no longer time reversal invariant.
In this paper we call Majorana nanowire (MNW) a semiconductor nanowire with
all three physical effects a), b) and c) mentioned above.

The transport properties in the presence of localized zero modes have been 
investigated for a normal-superconductor interface \cite{law09,fle10,wim11}.
It has been shown that both for resonant tunneling and for
transmission through a quantum point contact a conductance quantization 
at half-integer multiples 
of $4e^2/h$ is obtained when a zero mode is present at the interface. 
In this case the 
superconductor is called topological. 
In this manuscript we address a 
related although different geometry, the N/MNW/N structure where N 
refers to normal contacts in which the pairing
and Rashba interactions vanish. We show that the existence of a
zero mode is characterized by a unitary Andreev reflection, giving
a linear conductance of the N/MNW/N structure
equal to $e^2/h$. 
We also find that by increasing
the Zeeman coupling parallel to the wire a pronounced dip in the 
linear conductance appears due to the mixing between 
channels induced by the Rashba interaction.
Similarly to the N/superconductor case \cite{wim11}, 
the differential conductance has a peak
at zero bias in presence of the zero mode, that evolves to a dip
when the zero mode is absent. It is also worth stressing that 
the role of disorder of the N/MNW/N system 
has been studied in Ref.\ \cite{bee10}. 

Our main contribution in this work is the formalism of the 
coupled channel model (CCM) for the Bogliubov-deGennes (BdG) 
Hamiltonian. This formalism can be viewed as an alternative to 
the methods based on matching of plane waves, or on tight-binding
chains \cite{bee94,kit01}. It is particularly adapted to the description
of spatially smooth potentials and it gives insights on the 
role of the different physical mechanisms by means of the 
channel-channel couplings. We present numerical solutions of the CCM 
equations for a representative case of a 2D MNW based on InAs.
In support of our physical interpretations, we also present a
simplified effective model allowing a fully
analytical solution.

\section{The physical system}

\begin{figure}[t]
\centerline{\includegraphics[width=9cm,clip]{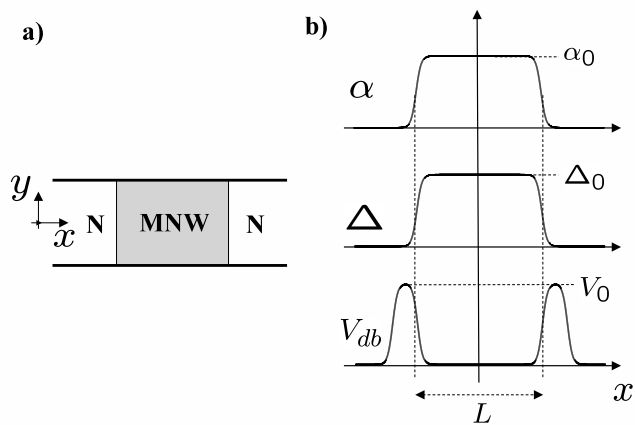}}
\caption{Sketch of the physical system. Panel a) displays our choice of 
coordinates while panel b) shows the longitudinal variation of the 
Hamiltonian parameters.}
\label{fig1}
\end{figure}

The N/MNW/N system is modelled as a
2D channel of transverse  dimension $L_y$ and with a 
central region of length $L$ with superconducting and Rashba interactions.  
These interactions vary smoothly in the longitudinal direction taking constant
values $\Delta_0$ and $\alpha_0$  in the MNW, and zero in the asymptotic 
regions of the leads.
In addition, potential barriers separate the central MNW from the leads.  
A sketch of the system and of the $x$-dependent functions is shown in Fig.\ \ref{fig1}.
The Hamiltonian reads
\begin{eqnarray}
\label{hbdg}
{\cal H}_{BdG}&=&(h_0-\mu)\tau_z
+\Delta(x)\tau_x+\Delta_B\vec{\sigma}\cdot\hat{n}\nonumber\\
&+& \frac{\alpha(x)}{\hbar}\left(p_x\sigma_y-p_y\sigma_x\right)\tau_z
+\frac{(p_x\alpha(x))}{2\hbar}\sigma_y\tau_z\; ,
\end{eqnarray}
with
\begin{equation}
h_0=
\frac{p_x^2}{2m^*}
+\frac{p_y^2}{2m^*}
+V_{db}(x)+V_c(y)\; .
\end{equation}
The $x$-dependence of the pairing $\Delta(x)$, of the Rashba coupling $\alpha(x)$
and of the double barrier $V_{db}(x)$ is modeled by smooth Fermi-like functions with a small diffusivity $d$; for instance 
\begin{equation}
\label{eqfer}
\Delta(x)=\Delta_0 \left(
\frac{1}{1+e^{(x-{\scriptsize\frac{1}{2}}L)/d}}
-
\frac{1}{1+e^{(x+{\scriptsize\frac{1}{2}}L)/d}}
\right)\; .
\end{equation}
The transverse confinement potential $V_c(y)$ is taken simply as an infinite square well
with zero potential at the bottom. The chemical potential 
explicitly appearing in the BdG theory is represented 
by parameter $\mu$ in Eq.\ (\ref{hbdg}); while 
$\vec\sigma$ and $\vec\tau$ are,
in a usual notation, the vectors of Pauli matrices acting in spin and isospin (or particle-hole) 
spaces, respectively.

A distinctive characteristic of our model is the continuity of the system parameters
with respect to the longitudinal coordinate $x$, as sketched in Fig.\ \ref{fig1}b. 
This would allow us to investigate, for instance, the dependence
on the diffusivity $d$ of the transition. In this work, however, we will assume rather steep 
transitions of the system parameters.
The coherent quasiparticle transport is described by the BdG equation
\begin{equation}
\label{bdgeq}
{\cal H}_{BdG}\Psi=E\Psi\; ,
\end{equation} 
where $E$ and $\Psi$ are the quasiparticle energy and wave function, respectively. The latter
depends on the position in space $(x,y)$ as well as on the spin and isospin 
variables $(\eta_\sigma,\eta_\tau)$, where $\eta=\uparrow,\downarrow$ represents a generic
discrete variable with only two possible values,
\begin{equation}
\Psi\Rightarrow\Psi(x,y,\eta_\sigma,\eta_\tau)\; .
\end{equation}
Notice, finally, that $\hat{n}$ is assumed to lie in the $xy$ plane. An out-of-plane component 
would require the addition of orbital magnetic effects not considered in this work.

\section{The coupled channel model}

We present in this section the description of transport in terms of channel amplitudes
or wave functions obeying a set of coupled differential equations. This description can be viewed
as an alternative to the matching of bulk solutions often used in the literature. The CCM is well
suited to the problem of a spatially continuous Hamiltonian posed in the preceding section. 
Before discussing the CCM equations, however, we need to consider the asymptotic solutions
($x\to\pm\infty$) as they are actually defining the channels themselves.

\subsection{Asymptotic solutions}

Since the pairing and Rashba intensities vanish asymptotically, the BdG Hamiltonian greatly simplifies
in those regions,
\begin{equation}
\lim_{x\to\pm\infty}{\cal H}_{BdG}=(h_0-\mu)\tau_z+\Delta_B\vec\sigma\cdot\hat{n}\; .
\end{equation}
In this limit the eigenstates are spinors pointing in the direction of $\hat{n}$ and 
$\hat{z}$ for spin and isospin. Introducing the quantum numbers $s_\sigma=\pm1$ and
$s_\tau=\pm1$ they read, respectively,
\begin{equation}
\chi_{s_\sigma}
\equiv
\frac{1}{\sqrt{2}}
\left(
\begin{array}{c}
1\\
s_\sigma e^{i\varphi}
\end{array}
\right) \quad ;\quad
\chi_{s_\tau}
\equiv\frac{1}{2}
\left(
\begin{array}{c}
1+s_\tau \\
1-s_\tau
\end{array}
\right)\;,
\end{equation}
where $\varphi$ is the azimuthal angle of $\hat{n}$. The spatial dependence of the asymptotic 
eigenstates is also analytical, a plane wave in $x$ and a square well eigenfunction in $y$,
$\phi_n(y)$, with $n=1,2,\dots$. Summarizing, a channel is specified by the quantum numbers $(ns_\sigma s_\tau)$
and its wave function reads 
\begin{equation}
\Psi_{ns_\sigma s_\tau}\equiv e^{ikx}\phi_n(y)\chi_{s_\sigma}(\eta_\sigma)
\chi_{s_\tau}(\eta_\tau)\; .
\end{equation}

The propagating or evanescent character of each channel is found when determining its 
wavenumber $k\equiv k_{ns_\sigma s_\tau}$. From Eq.\ (\ref{bdgeq}) 
the asymptotic BdG energy is
\begin{equation}
\label{EEE}
E=\left(  \frac{\hbar^2k_{ns_\sigma s_\tau}^2}{2m^*}+\varepsilon_n-\mu \right) s_\tau + \Delta_B s_\sigma\; ,
\end{equation}
where
\begin{equation}
\varepsilon_n= \frac{\pi^2 n^2\hbar^2}{2m^* L_y^2}\; .
\end{equation}
Inverting Eq.\ (\ref{EEE}) it is
\begin{equation}
\label{kkk}
k_{ns_\sigma s_\tau}=\sqrt{
\frac{2m^*}{\hbar^2}
\left(
E s_\tau-\varepsilon_n+\mu-\Delta_B s_\sigma s_\tau
\right)
}\; .
\end{equation}
The channel wavenumber $k_{ns_\sigma s_\tau}$ from Eq.\ (\ref{kkk}) is either real or purely imaginary. These two cases clearly correspond to propagating and evanescent channels, respectively.
In conclusion, for electrons ($s_\tau=1$) and holes ($s_\tau=-1$) the condition for propagating
mode of spin $s_\sigma$ in direction of $\hat{n}$ and with transverse state $n$ is 
\begin{equation}
\label{props}
E s_\tau-\varepsilon_n+\mu-\Delta_B s_\sigma s_\tau > 0\; .
\end{equation}

\subsection{The CCM equations}

 Assume the following expansion of the full wave function, valid not only in the asymptotic leads
 but for any arbitrary position,
 \begin{equation}
 \label{ccmwf}
 \Psi(x,y,\eta_\sigma,\eta_\tau) = \sum_{n s_\sigma s_\tau}{
 \psi_{n s_\sigma s_\tau}(x) \phi_n(y)
 \chi_{s_\sigma}(\eta_\sigma)
 \chi_{s_\tau}(\eta_\tau) }\; ,
 \end{equation}
 where $\psi_{n s_\sigma s_\tau}(x)$ is a 1D function we call the channel amplitude. Obviously, the channel amplitudes in the asymptotic regions are  $\psi_{n s_\sigma s_\tau}(x)\propto\exp{(\pm ik_{n s_\sigma s_\tau}x)}$,
 i.e., propagating or evanescent waves to the right or left directions, depending on the sign of the exponent. 
 The equations fulfilled by the channel amplitudes can be obtained 
 substituting the wave function, Eq.\ (\ref{ccmwf}), in the BdG equation,
 Eq.\ (\ref{bdgeq}), and projecting on a specific channel,
 \begin{equation}
 \label{proj}
\sum_{\eta_\sigma\eta_\tau}\int_0^{L_y}{dy}\,
\phi_n(y)
\chi^*_{s_\sigma}(\eta_\sigma)
\chi^*_{s_\tau}(\eta_\tau)
\;\times{
\left[ {\cal H}_{BdG}\Psi = E \Psi \right]
}\; .
\end{equation}

The sets of transverse wave functions $\{\phi_n\}$, $\{\chi_{s_\sigma}\}$ and 
$\{\chi_{s_\tau}\}$ fulfills proper orthonormality relations. After 
some straightforward algebra, Eq.\ (\ref{proj}) leads to 
\begin{eqnarray}
\label{ccm1}
\!\!\!\!\!\!\!\!\!\!\!\!\!\!\!\!\!\!\!\!
\left[
\left(\frac{p_x^2}{2m^*}+V_{db}(x)+\varepsilon_n-\mu\right) s_\tau
+\Delta_B s_\sigma
+ 
s_\sigma s_\tau\sin\varphi \frac{1}{2\hbar} 
\{p_x,\alpha(x)\}
-E \right]
\psi_{n s_\sigma s_\tau}(x)&&\nonumber\\
+ s_\sigma s_\tau\cos\varphi \frac{i}{2\hbar} 
\{p_x,\alpha(x)\}\, \psi_{n \bar{s_\sigma} s_\tau}(x) 
+\Delta(x)\, \psi_{n s_\sigma \bar{s_\tau}}(x) 
&&\nonumber\\
- s_\sigma s_\tau \frac{\alpha(x)}{\hbar}
\sum_{n'(\ne n)}{
\langle n | p_y | n'\rangle
\left[
\rule{0cm}{0.45cm}
\cos\varphi\, \psi_{n' s_\sigma s_\tau}(x)
-i
\sin\varphi\, \psi_{n' \bar{s_\sigma} s_\tau}(x)
\right]
}
= 0\; , &&
\end{eqnarray}
where we have introduced the usual anticommutator notation, 
$\{p_x,\alpha(x)\}=p_x\alpha(x)+\alpha(x)p_x$ and  
the bar over an index denotes its opposite value.
The set of Eqs.\ (\ref{ccm1}) is already a first version of our desired
CCM equations. There are three types of contributions to Eq.\ (\ref{ccm1}): 
a) the {\em background terms} of channel $(n s_\sigma s_\tau)$ are given
by the first line, b) the second line contains the coupling terms 
with channels of the same $n$ but
with opposite spin $\bar{s_\sigma}$ or isospin $\bar{s_\tau}$ to that 
of the background channel, c) finally, the third line shows the coupling 
with channels of a different $n$, the same isospin and 
arbitrary spin.

The physical role played by the different Hamiltonian contributions 
are clearly seen in Eq.\ (\ref{ccm1}). As expected, the superconducting 
pairing 
$\Delta(x)$ couples electron and hole channels. The two Rashba 
terms have a markedly different effect regarding the $n$ quantum number.
The $\alpha(x)p_x$ is diagonal in $n$, while the $\alpha(x)p_y$ is mixing
channels with different $n$'s with the selection rules imposed by 
the square well matrix element $\langle n|p_y|n'\rangle$. The relevance 
of the field orientation is also appreciated from Eq.\ (\ref{ccm1}).  For
instance, if the field is along $y$ ($\varphi=\pi/2$) the
mixing of $(n s_\sigma s_\tau)$ and $(n \bar{s_\sigma} s_\tau)$ vanishes.

We end this section mentioning a useful transformation of Eq.\ (\ref{ccm1})
that eliminates the linear terms in $p_x$ of the background problem.
Let us define the transformed channel amplitude
\begin{equation}
\label{gauge}
\tilde\psi_{n s_\sigma s_\tau} (x) =
e^{is_\sigma\sin\varphi\, {\cal K}_R(x)}
\psi_{n s_\sigma s_\tau} (x)\; , 
\end{equation}
where we introduced the dimensionless function
\begin{equation}
{\cal K}_R(x) = \frac{m^*}{\hbar^2}\int_0^x{dx' \alpha(x')}\; .
\end{equation}
Substituting Eq.\ (\ref{gauge}) into Eq.\ (\ref{ccm1}) we find 
\begin{eqnarray}
\label{ccm2}
&&
\!\!\!\!\!\!\!\!\!\!
\left[
\left(\frac{p_x^2}{2m^*}+V_{db}(x)+\varepsilon_n-\mu
-\frac{m^*}{2\hbar^2}\alpha(x)^2 \sin^2\varphi
\right) s_\tau
+\Delta_B s_\sigma
-E \right]
\tilde\psi_{n s_\sigma s_\tau}(x)\nonumber\\
&+& 
\left[
s_\tau \cos\varphi 
e^{2is_\sigma\sin\varphi\, {\cal K}_R(x)}  
\left(
i\frac{m^*}{\hbar^2}\alpha(x)^2 \sin\varphi
+
s_\sigma 
\frac{i}{2\hbar}\{p_x,\alpha(x)\}
\right)
\right]
\, \tilde\psi_{n \bar{s_\sigma} s_\tau}(x)\nonumber\\ 
&+&\Delta(x)\, \tilde\psi_{n s_\sigma \bar{s_\tau}}(x)
- s_\sigma s_\tau \frac{\alpha(x)}{\hbar}
\sum_{n'(\ne n)}{
\langle n | p_y | n'\rangle
\left[
\rule{0cm}{0.5cm}
\cos\varphi\, \tilde\psi_{n' s_\sigma s_\tau}(x)
\right.} \nonumber\\
&& \qquad\qquad\qquad\qquad\left.
\rule{0cm}{0.5cm}
-i
\sin\varphi e^{2is_\sigma\sin\varphi\, {\cal K}_R(x)}\, 
\tilde\psi_{n' \bar{s_\sigma} s_\tau}(x)
\right]
= 0\; ,
\end{eqnarray}

The set of Eqs.\ (\ref{ccm2}) is very similar to (\ref{ccm1}), with two
important differences: a) the background channel terms have a new
contribution quadratic in $\alpha(x)$ which is 
spin-independent, while the 
contribution linear in $p_x$ is effectively eliminated from this channel, 
b) the position-dependent phase
of the transformation given in Eq.\ (\ref{gauge}) appears
explicitly in the coupling with $(n \bar{s_\sigma} s_\tau)$
and $(n' \bar{s_\sigma} s_\tau)$.

\subsection{The QTBM}

We have solved the set of Eqs.\ (\ref{ccm1}) using the quantum-transmitting-boundary
formulation of the scattering problem. The reader is addressed to Refs.\ \cite{len90,san06} for 
details on the QTBM. Here we just mention for the sake of completeness the basic underlying ideas.
Using a 1D grid Eq.\ (\ref{ccm1}) can be discretized
with finite-difference formulas for the derivatives. In the asymptotic regions of the leads we impose
the analytical solutions of the channel amplitudes
\begin{equation}
\label{qtbm1}
\psi_{n s_\sigma s_\tau}(x)=
a^{(i)}_{n s_\sigma s_\tau} e^{i s_i s_\tau k_{n s_\sigma s_\tau} (x-x_i)}
+
b^{(i)}_{n s_\sigma s_\tau} e^{-i s_i s_\tau k_{n s_\sigma s_\tau} (x-x_i)}
\; ,
\end{equation}
where $a^{(i)}_{n s_\sigma s_\tau}$ and $b^{(i)}_{n s_\sigma s_\tau}$ are the usual
incident and reflected amplitudes in lead $i$. In Eq.\  (\ref{qtbm1}) we have introduced the 
{\em lead sign} $s_i$, equal to +1 and $-1$ for the left ($i=1$) and right ($i=2$)
leads, respectively, as well as the position of each lead boundary  $x_i$. 
We have also taken into account the 
reversed direction of propagation for electrons and holes with the $s_\tau$ sign.
Notice that from Eq. (\ref{qtbm1}) the outgoing coefficient $b^{(i)}_{n s_\sigma s_\tau}$ is 
expressed in terms of the channel amplitude at the lead boundary,
$b^{(i)}_{n s_\sigma s_\tau} = \psi_{n s_\sigma s_\tau}(x_i)-
a^{(i)}_{n s_\sigma s_\tau}$. Substituting this explicit expression of 
$b^{(i)}_{n s_\sigma s_\tau}$ back in Eq.\ (\ref{qtbm1}) we obtain
\begin{equation}
\label{qtbm2}
\!\!\!\!\!\!\!\!\!\!\!\!\!\!\!
\psi_{n s_\sigma s_\tau}(x)
-
e^{-i s_i s_\tau k_{n s_\sigma s_\tau} (x-x_i)}
\psi_{n s_\sigma s_\tau}(x_i)
= 2i\sin{\left(s_i s_\tau k_{n s_\sigma s_\tau} (x-x_i)\right)}
a^{(i)}_{n s_\sigma s_\tau}\\* \; .
\end{equation}

The QTBM closed system of linear equations is defined as follows: a) for a grid point $x$ such that 
$x_1\le x\le x_2$ we impose the discretized version of Eq.\ (\ref{ccm1}), b) for a grid point
having $x<x_1$ or $x>x_2$ we impose Eq.\ (\ref{qtbm2}). The resulting linear system
has as many equations as grid points and it 
depends only on the set of input coefficients
$\{a^{(i)}_{n s_\sigma s_\tau}\}$. It is highly 
sparse and can be numerically solved in an efficient way. 

The matrix of transmissions from mode 
$n s_\sigma s_\tau$ of lead $i$ to mode
$n' s'_\sigma s'_\tau$ of lead $i'$ is given by
\begin{equation}
\label{ttt}
t(i' n' s'_\sigma s'_\tau\leftarrow i n s_\sigma s_\tau)
=
\left.
\frac{
\sqrt{k_{n' s'_\sigma s'_\tau}}\, b^{(i')}_{n' s'_\sigma s'_\tau}
}{
\sqrt{k_{n s_\sigma s_\tau}}\, a^{(i)}_{n s_\sigma s_\tau}
}\right|_{\rm oim}\; ,
\end{equation}
where the subscript oim, standing for only incident mode, refers to the fact that all 
incident amplitudes vanish except the one explicitly appearing in the denominator
of Eq.\ (\ref{ttt}).
For use in the next section, we define a reduced matrix of transmission probabilities 
where we only discriminate lead and particle type,
\begin{equation}
\label{PPP}
P_{i'i}^{s'_\tau s_\tau} =
\sum_{nn' s_\sigma s'_\sigma}{
\left|\,t(i' n' s'_\sigma s'_\tau\leftarrow i n s_\sigma s_\tau)\,\right|^2
}\; .
\end{equation}

\section{Transport in the BdG framework}

The description of transport through MNW's can be done 
with the formalism of transport through normal/superconductor/normal
structures. We follow, specifically, the formulation by Lambert et al.\ 
\cite{lam93} for mesoscopic superconductors. For our two-terminal structure,
labelled as $i=1,2$ for left and right contacts, the current in terminal
$i$ reads
\begin{equation}
\label{eqI}
I_i=\int_0^\infty{
\sum_{\alpha=\pm1}{\alpha
\left(J_i^\alpha(E)-\hat{J}_i^\alpha(E)\right)
}\, dE
}\; ,
\end{equation}
where $E$ is the BdG quasiparticle energy. In Eq.\ (\ref{eqI}), 
$J_i^\alpha(E)$ and $\hat{J}_i^\alpha(E)$ are, respectively, the in-going and out-going fluxes
in lead $i$ of type $\alpha$. 

The essential ingredients we need to specify in order to 
use Eq.\ (\ref{eqI}) are the 
quasiparticle energy distributions $f_i^\alpha(E)$, the number of 
propagating modes $m_i^\alpha(E)$, and the matrix of quantum transmissions
$P_{ij}^{\alpha\beta}(E)$. The latter two are obtained from the CCM, 
Eqs. (\ref{props}) and (\ref{PPP}), respectively. The quasiparticle distributions are assumed to 
be given by Fermi functions
\begin{equation}
f_i^{\alpha}(E) = 
\left[
1+e^{\left(E-\alpha e V_i\right)/KT}
\right]^{-1}\; ,
\end{equation}
where the $i$-th reservoir chemical potential has been defined as
$\mu_i=\mu+eV_i$ and $kT$ is the thermal energy.
With these inputs the fluxes in Eq.\ (\ref{eqI}) read
\begin{eqnarray}
J_i^\alpha(E) &=& \frac{e}{h} m_i^\alpha(E) f_i^\alpha(E)\; ,\\
\hat{J}_i^{\alpha}(E) &=& \frac{e}{h} \sum_{j\beta} P_{ij}^{\alpha\beta}(E) f_j^\beta(E)\; .
\end{eqnarray}

This formalism fulfills two basic physical conditions:
a) vanishing of current for zero bias and b) equality
of current in both leads.
Indeed, for zero bias all distributions are 
identical $f_i^\alpha(E)\equiv f(E)$
and then the sum rule on quantum  transmissions,
\begin{equation}
\label{eqSR}
\sum_{j\beta}{P_{ij}^{\alpha\beta}(E)}=m_i^\alpha(E)\; ,
\end{equation}
ensures that in-going and out-going fluxes exactly cancel each other. 
The second condition, $I_1+I_2=0$, is more subtle; following Lambert \cite{lam93} we
interpret that it actually determines the MNW chemical potential $\mu$, relative to $\mu_1$ and
$\mu_2$. Notice that the potential bias between the two leads is $V=V_1-V_2$ and that the MNW chemical potential lies somewhere in the range between the two reservoir chemical potentials, 
\begin{equation}
\min(\mu_1,\mu_2) \le \mu\le \max(\mu_1,\mu_2)\; .
\end{equation}

The following practical approach to the BdG transport problem is then suggested: 
1) given $\mu_1$ and $\mu_2$,
assume $\mu=(\mu_1+\mu_2)/2$ and solve the BdG-CCM equations for the set 
$\{m_i^\alpha, P_{ij}^{\alpha\beta}\}$; 
2) compute $I_1+I_2$;
3) vary the value of $\mu$ 
and recompute $\{m_i^\alpha, P_{ij}^{\alpha\beta}\}$
until $I_1+I_2=0$ is fulfilled. Solving this selfconsistency loop might be a 
difficult task, however it is not needed when
the problem is symmetric with respect to $x$ inversion around the center of the MNW. 
In this case $\mu=(\mu_1+\mu_2)/2$ is already the solution giving $I_1+I_2=0$
since the bias $V$ has to be shared symmetrically, $V_i=s_iV/2$, where $s_1=1$ and $s_2=-1$.
Here we shall focus on the symmetric problem, leaving for a future work 
the analysis on the non symmetric case.

\subsection{Differential and linear conductances}

The differential conductance, defined generically as $dI/dV$, is 
one of the most relevant transport properties usually measured 
in experiments.
At zero temperature, the above formalism yields a very 
simple expression of this quantity because, in this limit, the 
derivatives of the quasiparticle distribution functions 
with respect to the bias
become Dirac deltas. Of course, this is true only in the symmetric case, when
$V=2s_iV_i$. 

For $T=0$ we obtain
\begin{equation}
\label{di1dv}
\frac{dI_1}{dV}=\frac{e^2}{2h}\left(
P_{12}^{++}({\scriptstyle\frac{1}{2}}eV)
+
P_{12}^{--}({\scriptstyle\frac{1}{2}}eV)
+
P_{11}^{+-}({\scriptstyle\frac{1}{2}}eV)
+
P_{11}^{-+}({\scriptstyle\frac{1}{2}}eV)
\right)\; ,
\end{equation}
and, as discussed above, it is $dI_2/dV=-dI_1/dV$.
The expression of the linear conductance $G$ can be obtained  
simply setting the bias to zero in Eq.\ (\ref{di1dv}). Using, in addition,
the particle hole symmetry
\begin{equation}
P_{ij}^{\alpha\beta}(E)=
P_{ij}^{\bar\alpha\bar\beta}(-E)\; ,
\end{equation}
we find 
\begin{equation}
\label{gg}
G=\frac{e^2}{h}\left(
P_{12}^{++}(0)
+
P_{11}^{+-}(0)
\right)
\; .
\end{equation}

Equations (\ref{di1dv}) and (\ref{gg}) are the basic relations of this work. Notice that 
they contain two qualitatively different contributions to the
conductance, a {\em normal transmission}, $T_0\equiv (P_{12}^{++}+P_{12}^{--})/2$, whereby 
quasiparticle type is conserved;  and 
an {\em Andreev reflection}, $R_A\equiv (P_{11}^{+-}+P_{11}^{-+})/2$, with 
quasiparticle change. Anticipating a result to be discussed below, notice
that Eqs.\ (\ref{di1dv}) and (\ref{gg}) predict a remarkable phenomenon, a
nonvanishing conductance in absence of transmission ($T_0=0$) due solely to 
Andreev reflection. This occurs when the Majorana nanowire has a 
zero mode.  In this case Andreev reflection is maximal for zero bias, while increasing the 
bias there is a reduction of $R_A$, i.e., a zero-bias
anomaly appears in $dI_1/dV$ due to the zero mode.

\section{Results and discussion}

\subsection{Physical and scaled values of the parameters}

The relative strengths of spin-orbit, pairing and Zeeman terms 
for a given transverse dimension $L_y$ are determined by the 
following scaled 
dimensionless ratios (scaling is indicated with an $s$ superscript) 
\begin{eqnarray}
\label{sc1}
\alpha_0^{(s)} &=& \frac{\alpha_0 m^*}{\hbar^2}L_y \;,\\
\label{sc2}
\Delta_0^{(s)} &=& \frac{\Delta_0 m^*}{\hbar^2}L_y^2 \; ,\\
\label{sc3}
\Delta_B^{(s)} &=& \frac{\Delta_B m^*}{\hbar^2}L_y^2 \; .
\end{eqnarray}
Notice that for a given set of physical values of $\alpha_0$, $\Delta_0$
and $\Delta_B$ different values of the transverse dimension $L_y$ will actually
correspond to different relative strengths through Eqs.\ (\ref{sc1}-\ref{sc3}).
Increasing $L_y$ the scenario clearly evolves from weak to strong couplings.

More specifically, we consider below physical parameters 
that could represent an InAs-based nanowire \cite{sch10}, 
$m^*=0.033 m_e$, $\alpha_0=30$ meVnm and a pairing gap of
$\Delta_0=0.3$ meV. We assume a realistic value of the wire transverse 
dimension, $L_y=150$ nm,
for which the relative strengths of Rashba and pairing are then
$\alpha_0^{(s)}\approx 2$ and $\Delta_0^{(s)}\approx 3$. 
Fixing these two scaled parameters to these values we will 
study the dependence on the third one $\Delta_B^{(s)}$ below.
The conversion
of the Zeeman coupling to a physical magnetic field is
$B=\Delta_B/g\mu_B$, with $g$ and $\mu_B$ the g factor and Bohr magneton,
respectively.
With 
our assumptions this conversion reads $B=(1.7 \Delta_B^{(s)}/g)$ T, in 
terms of the scaled Zeeman coupling.
That is, 
$B= 1$ T would correspond to $\Delta_B^{(s)}= 10$
for a g factor $\approx 17$.

The distance $L$ between barriers (Fig.\ \ref{fig1}) 
is taken as $L=20L_y=3$ $\mu{\rm m}$,
with barrier thickness of 150 nm and height $V_0=0.5$ meV, while
the spatial diffusivity is $d=15$ nm [see, e.g., Eq.\ (\ref{eqfer})].
We also choose the chemical potential, defining our reference energy
of the MNW, as $\mu=0$. 
Overall, we stress that 
the complete parameter set is representative of a typical experiment with an InAs-based
2D semiconductor wire.

\begin{figure}[t]
\centerline{\includegraphics[width=8cm,clip]{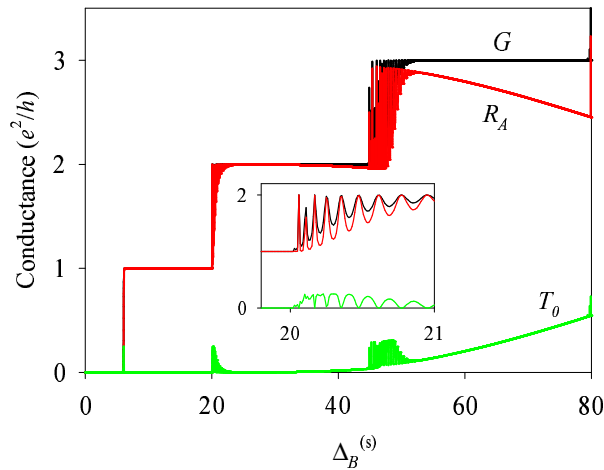}}
\caption{(Colour online) Linear conductance in absence of Rashba mixing
as a function of the scaled Zeeman energy.
The contributions from Andreev reflection  ($R_A$) and normal transmission ($T_0$)
 are also shown. 
 The inset shows a blow-up of the data in a small region. The system parameters are given in Subsection 5.1.
}
\label{fig2}
\end{figure}

\subsection{Linear conductance results}

\begin{figure}[t]
\centerline{\includegraphics[width=8cm,clip]{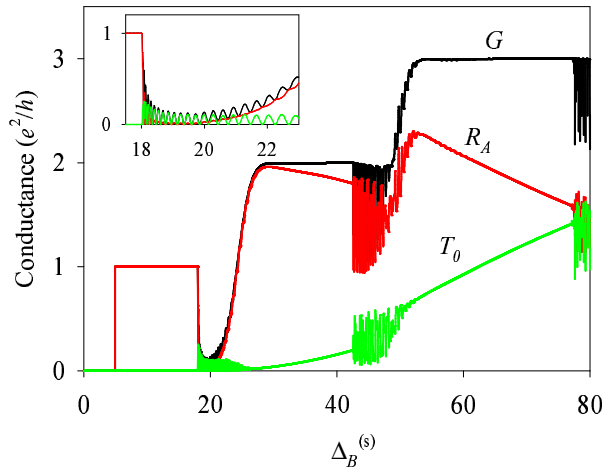}}
\caption{(Colour online) 
Same as Fig.\ \ref{fig2} including Rashba mixing.
}
\label{fig3}
\end{figure}

\begin{figure}[t]
\centerline{\includegraphics[width=8cm,clip]{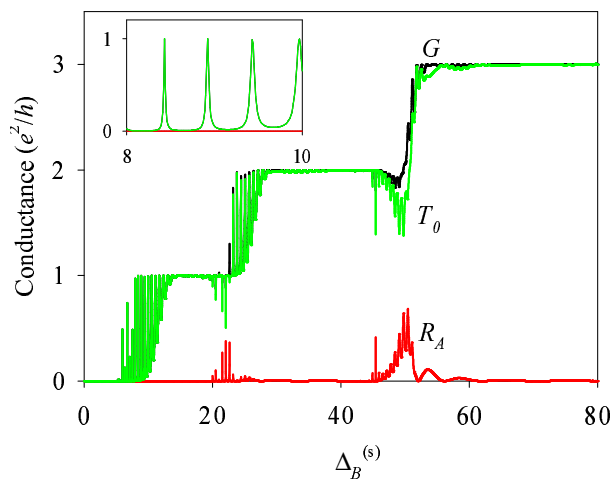}}
\caption{(Colour online) 
Same as Fig.\ \ref{fig3} for magnetic field along $y$.
}
\label{fig4}
\end{figure}

In Figs.\ \ref{fig2} to \ref{fig4}
we display the linear conductance calculated from  Eq.\ (\ref{gg}) 
as a function of the scaled Zeeman value $\Delta_B^{(s)}$ for the set of parameters 
mentioned in the preceding subsection. 
Figures \ref{fig2} and \ref{fig3} correspond to magnetic field in parallel direction to the wire ($x$)
while Fig.\ 4 is for transverse orientation ($y$).
In Figure 2 we neglected the contribution of the 
Rashba mixing, i.e., of the terms containing $\alpha(x)p_y$ in Eq.\ (\ref{ccm1}). 
Notice that in this situation the linear conductance displays an almost perfect quantization
in  $e^2/h$ steps. Increasing $\Delta_B^{(s)}$, small deviations in the form of
very narrow spikes  can be seen at the beginning of the second and third plateaus. In the first two 
steps all the conductance is due to Andreev reflection since the normal transmission 
is negligible. Perfect Andreev reflection is a signal of the existence of a zero mode
of the closed system, i.e., a
Majorana fermion bound at the interface between the MNW and the normal contacts.
This zero mode yields a perfectly quantized conductance in absence of transmission
due solely to Andreev reflection, i.e., $G=(e^2/h) R_A$.
The decrease of Andreev reflection for $\Delta_B^{(s)}>50$
in Fig.\ \ref{fig2} can be attributed to the finite size effect that removes the Majorana modes
from perfect zero energy, in agreement with the analysis of Ref.\ \cite{PRBrc} for the closed system.
This decrease in $R_A$ is accompanied by an increase in $T_0$, keeping the value of 
$G$ close to integer multiples of $e^2/h$, except at the transition between steps.

Figure \ref{fig3} displays the linear conductance for the same system of Fig.\ \ref{fig2}, but
now including the full Rashba interaction. A conspicuous difference with Fig.\ \ref{fig2} is that 
the conductance deviates from the simple staircase behaviour, with a broad conductance dip 
appearing at the end of the first plateau. This dip is due to a magnetic instability 
precluding the formation of two zero modes
due to a repulsion between modes induced by Rashba mixing \cite{PRBrc}.
The effect of this mechanism on the linear conductance is remarkable, with
the prediction  of a reduced conductance due to
a large reduction of Andreev reflection. This anomalous behaviour
of the conductance at the end of the conductance plateau also appears in higher 
plateaus, as seen in Fig.\ \ref{fig3} for the second  and third plateaus. Notice, however, that the 
finite size effect mentioned above transforms the conductance dips
of the higher plateaus in a strongly oscillating
behaviour. We have checked that the formation of the conductance dips due to the 
instability of multiple zero modes is even more robust with 
higher values of the pairing gap and Rashba strengths, and that it is also robust against
variations of the barriers between the normal contacts and the MNW.

Figure \ref{fig4} shows the evolution of the linear conductance with
$\Delta_B^{(s)}$ for field along the transverse direction $y$. For this
orientation of the field the physics changes completely, since now
it is the Andreev reflection that vanishes and the conductance is due 
to the normal transmission. Only small peaks in $R_A$ can be 
seen in the transition between plateaus.
The vanishing of $R_A$ is due to the absence of zero modes 
of the closed MNW 
for magnetic fields along $y$ \cite{PRBrc}. A similar orientation
anisotropy has 
been seen in experiments with cylindric InSb nanowires \cite{mou12,den12}.
No conductance dips are observed in Fig.\ \ref{fig4} but
there are many spikes due to resonant transmission
through the double-barrier potential $V_{db}(x)$. 
The separation
between spikes is very small due to the
dense distribution of quasibound states for such a
long system $L=3$ $\mu$m. From this point of view, it is still 
more remarkable 
the fact that for field along $x$ the presence of a zero mode
washes the spike oscillations and yields a consistent maximal 
conductance in some regimes.

\subsection{Nonlinear conductance}

\begin{figure}[t]
\centerline{\includegraphics[width=7cm,clip]{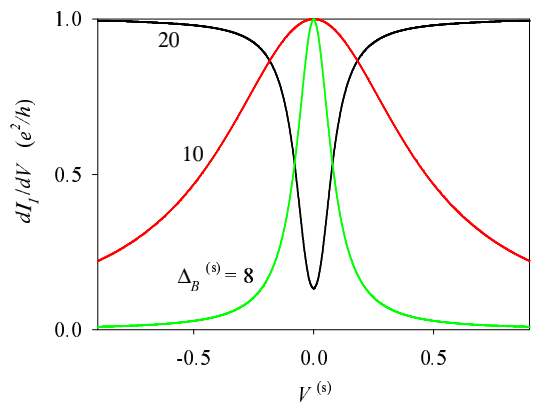}}
\caption{(Colour online) 
$dI_1/dV$ as a function of scaled bias $V^{(s)}$ for the same parameters of
Fig.\ \ref{fig3} and for different values of $\Delta_B^{(s)}$.
}
\label{fig5}
\end{figure}

The nonlinear conductance obtained with Eq.\ (\ref{di1dv}) 
is shown in Fig.\ \ref{fig5} as a function of the applied bias.
We have taken some selected values of $\Delta_B^{(s)}$ from
Fig.\ \ref{fig3}, corresponding to vanishing bias, 
and explored the variation with $V$. As in the preceding subsection 
we define a scaled bias taking the transverse confinement as
reference, i.e., 
$V^{(s)}=(em^*L_y^2/\hbar^2)V$.
For $\Delta_B^{(s)}=9$
there is a narrow peak in $dI_1/dV$ at zero bias. This 
zero bias anomaly is reflecting the existence of a zero mode in the MNW.
Increasing the Zeeman coupling the peak broadens, becoming a 
flat distribution. For $\Delta_B^{(s)}=19$,  corresponding to
the conductance dip of Fig.\ \ref{fig3}, the zero bias peak 
changes to a zero bias minimum. The existence of a zero bias
anomaly in the presence of zero modes has been 
discussed before in systems with a superconductor contact, 
experimentally
in Refs.\ \cite{mou12,den12}
and
theoretically in Refs.\ \cite{law09,fle10,wim11}. Our results 
prove that a similar behaviour is to be expected in N/MNW/N
structures. 

\subsection{Density distributions}

\begin{figure}[t]
\centerline{\includegraphics[width=8cm,clip]{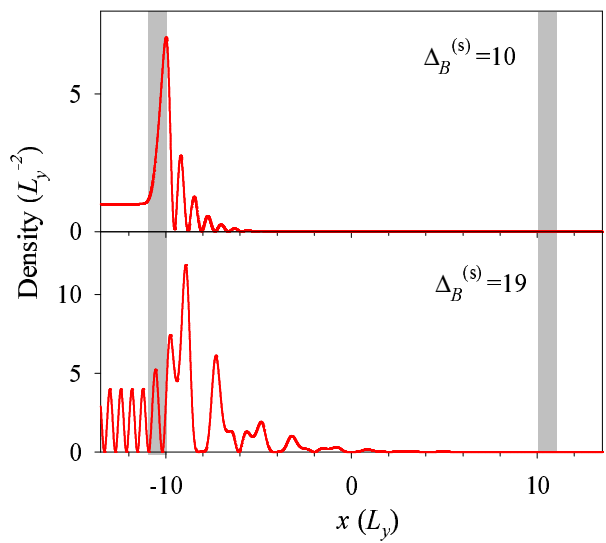}}
\caption{ 
Density distributions $|\psi_{ns_\sigma s_\tau}(x)|^2$ 
for two values of $\Delta_B^{(s)}$ 
of the first plateau of Fig.\ \ref{fig3}.
We assumed boundary conditions corresponding to incidence
from the left side. For comparison, the position of the 
potential barriers is indicated by the shaded regions.
}
\label{fig6}
\end{figure}

The density distributions, defined as
$|\psi_{ns_\sigma s_\tau}(x)|^2$, are shown in Fig.\ \ref{fig6}
for two values of $\Delta_B^{(s)}$. They correspond to the perfect
Andreev reflection ($\Delta_B^{(s)}=10$) and
to the conductance dip ($\Delta_B^{(s)}=19$) of Fig.\ \ref{fig3}.
As expected, the upper panel shows that the incident
unitary density couples with an edge mode of the MNW.
The density profile localized at the edge and decaying towards
the interior has exactly the same shape found 
in calculations of zero modes of closed MNW's \cite{PRBrc}. Perfect Andreev reflection
in this situation consists in the total reflection in the
conjugate channel and, therefore, no quantum interference is observed
in the left contact. As mentioned before, this occurs due to the 
presence of the zero mode in the MNW and it allows unit
conductance without any transmission at all between left and 
right contacts. 

The lower panel of Fig.\ \ref{fig6} shows a qualitatively different
behaviour. The beating pattern in the left contact is indicating
that full reflection occurs now in the same channel of incidence,
with a strong interference between incident and reflected waves 
The density at the edge of the MNW is more irregular and extends 
farther towards the interior than in the upper panel. 
The physical interpretation is clear:
for this Zeeman intensity the edge 
mode of the MNW lies at a nonzero energy, this causing
normal reflection, as opposed to the Andreev reflection
of the upper panel.

\section{A model in second quantization}

To provide additional insight on the physics of
transport through the MNW, 
in this section we consider an effective Hamiltonian in second quantization projected onto the Majorana subspace.
The Hamiltonian consists of three parts, that is,
\beq
\calh_{\it eff} = \calh_C + \calh_M + \calh_T \; ,
\edq
with
\numparts
\begin{eqnarray}
\calh_C &=& \sum_{\alpha=L/R,k} \veps_{\alpha k} c_{\alpha k}^{\dag} c_{\alpha k} \; , \\
\calh_M &=& \frac{i}{2}\veps_M \eta_L \eta_R \; , \\
\calh_T &=& \sum_{\alpha,\beta,k} \left(V_{\alpha k,\beta}^{\ast} c_{\alpha k}^{\dag}\eta_{\beta} + V_{\alpha k,\beta}\eta_{\beta}c_{\alpha k}\right) \; .
\end{eqnarray}
\endnumparts
Here $\calh_C$ describes the normal leads, 
with $c_{\alpha k}^{\dag}$ ($c_{\alpha k}$) the Dirac fermion creation (annihilation) operator. 
Notice that the spin degree of freedom is neglected.
This can be understood considering that we need to apply a large magnetic field to observe the edge Majoranas, 
so that only one kind of spin is effectively involved \cite{Fle11}.
$\calh_M$ characterizes the coupled Majorana states, with $\eta_{L/R}$ Majorana fermion operators
fulfilling 
$\eta_i = \eta_i^{\dag}$, $\eta_i^2 = 1$ and with anticommutator relation $\{\eta_i,\eta_j\} = 2\delta_{ij}$. 
The parameter $\veps_M$ denotes the coupling between the two Majoranas 
on opposite ends of the MNW and can be some complicated function of wire length, superconducting coherence length, 
applied magnetic field, Rashba coupling and superconducting gap.
$\veps_M$ might be found by exact diagonalization of Hamiltonian (\ref{hbdg})
\cite{PRBrc}.
We will assume that it is known for the purpose of the present model.
The last contribution, $\calh_T$, 
corresponds to the tunnel Hamiltonian between normal leads and the Majoranas on opposite ends 
\cite{fle10}.
Below, the tunnel amplitude $V_{\alpha k,\beta}$ is taken as $V$ for $\alpha=\beta$ and zero for $\alpha \ne \beta$.

The current is computed from
\beq
I_{\alpha} = -\frac{ie}{\hbar} \langle[\calh,n_{\alpha}]\rangle
= -\frac{2e}{\hbar} \Re\left\{ \sum_{\beta} \sum_{k\in\alpha} V_{\alpha k,\beta} \calg_{\alpha k,\beta}^<(t,t)\right\} \; ,
\edq
where $n_{\alpha} = \sum_{k\in\alpha} c_{\alpha k}^{\dag}c_{\alpha k}$ and 
$\calg_{\alpha k,\beta}^<$ denotes the lesser component of the mixed Green's function defined as
\beq
\calg_{\alpha k,\beta}(t,t') = -i\langle T_K c_{\alpha k}(t)\eta_{\beta}(t')\rangle \; .
\edq
Employing the equation of motion technique, after tedious algebra, the current becomes
\beq
I_{\alpha} = -\frac{2e}{\hbar} \Im \left\{ \int d\veps~ 
\Tr\left[f_{\alpha}(\veps)\left(\bcalg_{\eta}^r(\veps)-\bcalg_{\eta}^a(\veps)\right)\bGamma_{\alpha}(\veps) + \bcalg_{\eta}^<(\veps)\bGamma_{\alpha}(\veps)\right]\right\} \; ,
\label{eq:Ialpha}
\edq
where $\bcalg_{\eta}(\veps)$ is the Green's function for the MNW and
$\bGamma_{\alpha}(\veps)$ denotes the hybridization matrix given by
\beq
\bGamma_{\alpha;\beta\gamma}(\veps) = \pi\sum_{k\in\alpha} V_{\alpha k,\beta}V_{\alpha k,\gamma}^{\ast}\delta(\veps-\veps_{\alpha k})
= \delta_{\alpha,\beta}\delta_{\beta,\gamma} \Gamma/2 \; .
\label{eq:Gamma}
\edq
To complete the calculation we need to determine the MNW Green's functions. They read \cite{fle10}
\begin{eqnarray}
\bcalg_{\eta}^{r/a}(\veps) &= \frac{2}{\veps - i\bt - 2\bSigma_{0}^{r/a}(\veps)} \; , \\
\bcalg_{\eta}^{<}(\veps) &= \bcalg_{\eta}^{r}(\veps) \bSigma_{0}^{<}(\veps)\bcalg_{\eta}^{a}(\veps) \; .
\label{eq:MNWG}
\end{eqnarray}
Here,
\beq
\bt = \left( \begin{array}{cc}
0        & \veps_M \\
-\veps_M & 0
\end{array} \right) \; ,
\label{eq:bt}
\edq
and the self-energy matrices are given by
\begin{eqnarray}
\bSigma_{0;\alpha\beta}^{r/a}(\veps) &= \mp i \sum_{\gamma} \left[\bGamma_{\gamma;\alpha\beta}(\veps)+\bGamma_{\gamma;\beta\alpha}(-\veps)\right] \; , \\
\bSigma_{0;\alpha\beta}^{<}(\veps) &= 2i \sum_{\gamma} \left[f_{\gamma}(\veps)\bGamma_{\gamma;\alpha\beta}(\veps)+f_{\gamma}(-\veps)\bGamma_{\gamma;\beta\alpha}(-\veps)\right] \; .
\label{eq:bSigma}
\end{eqnarray}
Substituting Eqs.\ (\ref{eq:Gamma})-(\ref{eq:bSigma}) into Eq.\ (\ref{eq:Ialpha}) and using current conservation we obtain
\begin{equation}
\!\!\!\!\!\!\!\!\!\!\!\!\!\!\!\!\!\!\!\!\
I_L = -I_R = \frac{e}{h} \int d\veps~\frac{4\Gamma^2\left(\veps^2 + 4\Gamma^2 + \veps_M^2\right)}{\left(\veps^2+4\Gamma^2\right)^2 + \veps_M^2\left(\veps_M^2-2\left(\veps^2-4\gamma^2\right)\right)}\left(f_L(\veps)-f_R(\veps)\right) \; .
\label{eq:Ifinal}
\end{equation}

From Eq.\  (\ref{eq:Ifinal}) we note that at $T=0$  the linear conductance finally reads
\beq
G = \frac{e^2}{h} \frac{4\Gamma^2}{\veps_M^2+4\Gamma^2} \; .
\edq
For zero energy Majoranas  $\veps_M = 0$ and then 
Eq.\  (\ref{eq:Ifinal})  yields $G = e^2/h$. This result 
nicely agrees with our interpretation of Fig.\ \ref{fig3} 
attributing 
maximal conductance to the zero mode. 

\section{Conclusions}

We have presented the formalism of transport in a N/MNW/N structure based on 
the coupled channel model. This formalism yields a transparent interpretation 
of the coupling between channels induced by the relevant physical mechanisms
of the problem.
Namely, the confinement, Zeeman, Rashba and superconducting interactions.
We have considered a 2D structure and in-plane magnetic fields, 
although the formalism can be extended
to consider more spatial dimensions
and different geometries.  

The coupled-channel-model equations have been solved 
using the quantum-transmitting-boundary algorithm for a set
of parameters representative of an InAs nanowire. The existence
of a zero mode in the MNW is characterized by a perfect Andreev
reflection, whereby an incident channel is totally reflected 
in its antiparticle conjugate one. For a single zero mode 
the linear conductance takes the maximal value $e^2/h$ due 
solely to Andreev reflection, without any quantum transmission 
from left to right contacts. For increasing values of the 
Zeeman coupling along the wire, a conspicuous dip in the linear 
conductance is predicted due to repulsion between Majoranas. This
repulsion originates in the Rashba mixing between channels.
On the contrary, for Zeeman coupling along $y$ the Andreev reflection vanishes,
with the possible exception of a small region close to 
the transition between plateaus. When the zero mode is absent, the
linear conductance has narrow spikes as a function of the 
Zeeman coupling. 

The differential conductance signals the presence of the 
zero mode with a peak at zero bias. The zero bias peak evolves to a 
dip when the MNW zero mode is absent. Finally, we have also discussed 
an effective model in second quantization confirming the physical interpretation
in terms of Majorana modes.
The coupled channel model presented in this work can be used to
investigate other scenarios like, e.g., non-symmetric barriers or 
sequential MNW's. Work along these lines is in progress.

\ack
We thank D. S\'{a}nchez for 
useful discussions. This work 
was supported by Grants 
No.\ FIS2008-00781, 
FIS2011-23526 and CSD2007-00042 (CPAN) 
of the Spanish Government.

\end{document}